\documentclass{svmult}
\usepackage{makeidx}         
\usepackage{graphicx}        
\usepackage{multicol}        
\usepackage[bottom]{footmisc}
\makeindex             

\usepackage{amsmath,url,booktabs}
\newcommand{\T}{^{\mathrm{T}}}		
\newcommand{\vek}[1]{\boldsymbol{#1}}	
\newcommand{\mat}[1]{\mathsf{#1}}	
\newcommand{\abs}[1]{\lvert #1\rvert}	
\newcommand{\refeq}[1]{Eq.~(\ref{#1})}	
\newcommand{\onefig}[1]{\includegraphics[scale=1.1]{#1}}	

\begin{document}
\title*{Network-based information filtering algorithms: ranking and recommendation}
\author{Mat\'u\v s Medo}
\institute{Physics Department, University of Fribourg, CH-1700 Fribourg, Switzerland;
\texttt{matus.medo@unifr.ch}}
\maketitle

After the Internet and the World Wide Web have become popular and widely-available, the electronically stored online interactions of individuals have fast emerged as a challenge for researchers and, perhaps even faster, as a source of valuable information for entrepreneurs. We now have detailed records of informal friendship relations in social networks, purchases on e-commerce sites, various sorts of information being sent from one user to another, online collections of web bookmarks, and many other data sets that allow us to pose questions that are of interest from both academical and commercial point of view. For example, which other users of a social network you might want to be friend with? Which other items you might be interested to purchase? Who are the most influential users in a network? Which web page you might want to visit next? All these questions are not only interesting per se but the answers to them may help entrepreneurs provide better service to their customers and, ultimately, increase their profits.

All of the questions posed above have many different ways to be approached that belong to the field of information filtering~\cite{HSS01}. The goal of information filtering is to eliminate the redundant or unsuitable information and thus overcome the information overload. In our case, information filtering helps users choose from an abundant number of possibilities (available products, potential friends, etc.) those that are most likely to be of interest or use for them. Common approaches to this task are based on mathematical statistics, machine learning, and artificial intelligence \cite{HTF09,WFH11}. They formulate a parametric mathematical model which is calibrated using the readily available data and then used to predict unknown user opinions.

In this chapter we discuss a different class of algorithms that all make use of a network representation of the data. The current classical example of such an algorithm is PageRank which, while having a far-reaching history \cite{Fran11}, has been re-invented and popularized by the founders of Google where it serves up to now as the key element of their Internet search engine \cite{BP98}. As we shall see below, this algorithm is closely related to random walks that play an important role in physics. (In the case of PageRank, of course, we do not face a random walk in physical space but a random walk on a network consisting of web pages and directed links among them.) These network-based methods can be used alone or in combination with other information filtering techniques, giving rise to hybrid methods \cite{Burke07}.

We focus here on two important information filtering tasks---ranking and recommendation. By ranking we mean producing a general list of available items (users or objects) that captures some inherent quality of theirs. Finding influential users or exceptional web pages belongs here. By recommendation we mean preparing a specific ``recommendation list'' for each individual user, each list containing items that are likely to be appreciated by the given user. Finding potential friends or items to purchase belong here. In addition to traditional unipartite networks where only nodes of one kind are present (such as the network of web sites connected by hyperlinks or a network of users connected by friendship relations), we will often make use of bipartite networks where nodes of two kinds are present. For example, a network connecting users with the items that they have purchased is bipartite because every link connects a user with an item while links between users or between items are entirely absent. For a review of networks and network analysis that do not directly contribute to ranking and recommendation yet they can help to understand the structure of the data at hand, see the survey of complex networks measurements in~\cite{CRTVB07}. For a general overview of dynamical processes on complex networks see~\cite{BBV08}.

\section{Ranking}
\label{sec:ranking}
When we want to rank nodes of the network, there are obviously many approaches, each of them suiting a different purpose. The simplest possible ranking is by node degree (or, in the case of a directed network, node in-degree) which is based on the assumption that ``important'' nodes are those that are referred by many other nodes. Many other measures of node importance exist, based either on local or global properties of the given network~\cite{Boccaletti2006}. In this section, we discuss importance rankings where score of a node is directly computed by random walk or where score spreads among the nodes in a manner akin to random walk.

\subsection{PageRank}
\label{sec:PR}
When given a directed unipartite network, PageRank~\cite{BP98} is arguably the most famous method to produce a general ranking of the network's nodes. The method is based on the circular idea \emph{``A node is important if it is pointed by other important nodes''} which can be applied to many different situations, including ranking of web sites (an important site is referred by important sites), scientific journals (articles from an important journal are cited by articles from important journals), and people (an important person is referred/trusted by important people). For a review of past research in this direction and the use of this circular idea in various disciplines see~\cite{Fran11}.

We begin with a general exposition of the approach, denoting the importance/score of node $i$ as $h_i$ and the non-negative strength of the link pointing from node $i$ to node $j$ as $w_{ij}$ ($i=1,\dots,N$ where $N$ is the number of nodes in the network). The above circular thesis can now be formalized as
\begin{equation}
\label{circular}
h_j=\sum_i \frac{w_{ij}}{\sum_k w_{ik}}\, h_i
\end{equation}
where the division with $\sum_k w_{ik}$ ensures that the importance of node $i$ is distributed among the nodes pointed by it with each node receiving part proportional to $w_{ij}$. To simplify our notation, we introduce normalized weights $P_{ij}:=w_{ij}/\sum_j w_{ij}$. Now we can write $h_j=\sum_i P_{ij}h_i$ which can be further simplified by matrix notation to get
\begin{equation}
\label{circular-matrix}
\vek{h}=\mat{P}\T\vek{h}.
\end{equation}
This matrix form shows that the sought-for vector $\vek{h}$ is the right eigenvector of $\mat{P}\T$ associated with eigenvalue $1$. Since $\mat{P}\T$ is now a column-normalized matrix (also called stochastic matrix), the Frobenius-Perron theorem applies and states that $1$ is its largest eigenvalue. A solution to \refeq{circular-matrix} thus always exists and when matrix $\mat{P}$ is irreducible, this solution is unique. (A matrix is irreducible if and only if in the directed graph that the matrix represents there exists a directed path between any two vertices.) The uniqueness is of course up to multiplication of $\vek{h}$ by a constant factor which allows us to impose the normalization condition $\sum_i h_i=1$. Note that \refeq{circular-matrix} is similar to that for eigenvector centrality measures that are common in the analysis of social networks~\cite{WaFa94,BoLo01}. In that case, however, one does not employ a normalized matrix $\mat{P}$ but the network's adjacency matrix itself and searches for a vector $\vek{x}$ satisfying $\mat{A}\T\vek{x}=\lambda\vek{x}$ where $\lambda$ is a number.

In addition to the redistribution point of view described above, a random walk view can often provide useful insights. The normalized weights $P_{ij}$ can be interpreted as probabilities of moving from node $i$ to node $j$ and, consequently, $h_i$ as the probability of being at node $i$. An initial probability distribution $\vek{h}^{(0)}$ transforms gradually by $\vek{h}^{(n+1)}=\mat{P}\T\vek{h}^{(n)}$ until a stationary probability distribution corresponding to the largest eigenvalue of $\mat{P}\T$ is established. If this eigenvalue is degenerated, the stationary solution is not unique. The rate of convergence of this iterative method is determined by the magnitude of the second-largest eigenvalue of $\mat{P}\T$.

Our treatment up to now was fully general and applies to any redistribution of $h_i$ values over a weighted network given by weights $w_{ij}$. Depending on the nature of the problem and the input data, one needs to choose the weights so that the resulting importance vector $\vek{h}$ contains the information that we are interested in. In the case of PageRank, which was designed to produce the importance score for web sites, the input data consists of a directed network of web sites where a hyperlink from site A to site B can be interpreted as a sign of approval of site B by site A. Since no additional strength information is attached to hyperlinks, the network of hyperlinks is represented by its adjacency matrix $\mat{A}$ where $A_{ij}=1$ if there is a link pointing from node $i$ to node $j$ (the network is directed and hence this matrix is not symmetric in general). Weights $P_{ij}$ thus should be the same for all sites $j$ referred by a given node $i$ which, respecting the weight normalization condition, leads to $P_{ij}=A_{ij}/k_i^o$ where $k_i^o$ is the out-degree of node $i$. Since this is ill-defined for nodes with no out-going links (``dangling nodes''), one usually assumes that if $k_i^o=0$, $P_{ij}=1/N$ for all $j$.

\begin{figure}[t]
\centering
\onefig{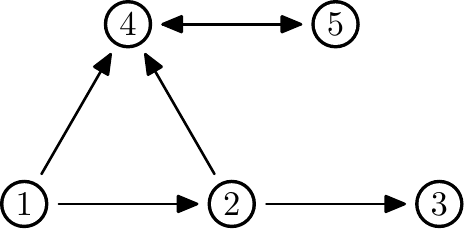}
\caption{PageRank computation for a toy network. When $\alpha=1$ (no random teleportation), scores of nodes 1, 2, and 3 goes with iterations to zero but no stationary distribution exist because one of the eigenvalues is $-1$ and causes ceaseless alternations of score of nodes 4 and 5. When $\alpha=0.85$ (the usual value adopted for web site ranking), the resulting score vector is $\vek{h}=(0.04,0.06,0.07,0.43,0.40)$. When $\alpha=0$ (no link-following), all nodes have the same score $1/5$.}
\label{fig:network}
\end{figure}

One can easily see that even when the problem of nodes with zero out-degree is solved, the resulting solution can easily be pathological in some sense. If the network contains a component without out-going links (so-called bucket, see nodes 4 and 5 in Fig.~\ref{fig:network}), this part of the network would act as a trap for the random walk process. The final solution $\vek{h}$ would concentrate there and thus it would give us little useful information. The inventors of PageRank overcame the problem by postulating that links leading from a node are followed only with certain probability $\alpha$~\cite{BP98}. With the complementary probability $1-\alpha$, teleportation (jump) occurs, ending at a randomly chosen node of the network. The corresponding transition matrix (also called Google matrix) is
\begin{equation}
\label{Gmatrix}
\mat{G}=\alpha\mat{P}\T+(1-\alpha)\mat{T}
\end{equation}
where $\mat{T}$ is the teleportation matrix with all elements equal $1/N$. The parameter $\alpha$ (also called damping), determines the weight given to link-following and teleportation, respectively. Since $\alpha$ is the probability of following an out-going link, one can easily compute that the average number of links followed in a row is
$\sum_{k=0}^{\infty} k\alpha^k(1-\alpha)=\alpha/(1-\alpha)$. In the original PageRank paper, $\alpha$ was proposed to be set around $0.85$ which corresponds to following five or six hyperlinks in a row and then jumping to a random page~\cite{BP98}. The value of $\alpha$ is closely related to the convergence rate of the iterative PageRank computation (the lower the value, the faster the convergence; see \cite{Berkhin05} for more details). While PageRank was originally devised for directed networks, one can apply it also to undirected networks~\cite{PeFo08,DYFC09}. The teleportation parameter then plays a crucial role---without it, PageRank score on an undirected network reduces to node degree.

Alternatively, one can replace the uniform teleportation matrix with $\vek{1}_N\vek{v}\T$ where $\vek{1}_N$ is an $N$-dimensional vector of ones and $\vek{v}$ is a normalized $N$-dimensional vector which allows us to give preference to some nodes. This provides an important additional degree of freedom and allows one to, for example, devise a topic-specific ranking as described in~\cite{HJSS06}. A complementary point of view is presented, for example, in~\cite{AgChAg06} where an inverse problem of finding matrix elements $G_{ij}$ from some partial knowledge of node-pair preferences (``we want the score of node $i$ to be higher than that of node $j$'') is studied.

Using the definition of $\mat{G}$ given in \refeq{Gmatrix}, the PageRank equation $\mat{G}\vek{h}=\vek{h}$ can be written as $\alpha\mat{P}\T\vek{h}+(1-\alpha)\vek{1}_N/N=\vek{h}$, leading to
\begin{equation}
\label{PageRank-analytical}
\vek{h}=\frac{1-\alpha}{N}\,(\mat{I}_N-\alpha\mat{P}\T)^{-1}\vek{1}_N=
\frac{1-\alpha}{N}\sum_{k=0}^{\infty} (\alpha\mat{P}\T)^k \vek{1}_N
\end{equation}
where $\mat{I}_N$ is an $N\times N$ identity matrix and $\vek{1}_N$ and $\vek{1}_N$ is an $N$-dimensional vector of ones. Here both the inverse and the series expansion exist as long as $\alpha<1$. While these formulas for computing $\vek{h}$ can be easily applied for small systems, a critical advantage of PageRank lies in the fact that the above-mentioned iterative method for finding $\vek{h}$ is in practice very effective even for very large systems. Thanks to that, PageRank serves as an important input for the Google's ranking of web sites where scores are computed for several billions of pages (for more information on the data mining for the WWW, see~\cite{LaMe06,Bing07}). Even for the enormous size of the WWW, only a few tens of iterations are sufficient to compute PageRank to a required precision~\cite{Havel99}. The iterative method is also easy to parallelize and, in addition, one can write $\vek{h}^{(n+1)}=\alpha\mat{P}\T\vek{h}^{(n)}+(1-\alpha)\vek{1}_N/N$ and thus benefit from the sparsity of $\mat{P}$. In comparison with that, directly multiplying $\mat{G}\vek{h}^{(n)}$ is computationally much more expensive because $\mat{G}$ has no zero entries.

Another advantage of PageRank is that it is robust to spamming and malicious behavior. This robustness is rooted in the inability of web site administrators to create new hyperlinks pointing to their sites. If they simply create fake new web sites pointing to the site whose status they want to enhance, the artificially created web sites themselves have low scores (because no one points at them) and contribute little to the score of the target site. Of course, various sophisticated methods of manipulating the PageRank still exist~\cite{CheF06}. The stability of node rankings obtained with PageRank is the central point in~\cite{GhBa11} where the authors show that PageRank is particularly prone to noisy data when the network is random (and thus the degree distribution, which is crucial for the ranking's stability, decays exponentially). By contrast, a small number of super-stable nodes whose ranking is particularly resistant to perturbations emerge in scale-free networks.

\subsection{Variants of PageRank}
From the conceptual point of view, an interesting generalization of PageRank has been proposed in~\cite{BYBC06} where spreading of the score was separated into branching (due to out-degree) and damping (due to the damping parameter $\alpha$). In the case of PageRank, damping is exponential because with each propagation step, another multiplication with $\alpha$ is added. The authors show that a power-law damping of the form $1/[(t+1)(t+2)]$ where $t$ denotes the number of steps is equivalent to a so-called TotalRank which is obtained simply by integrating the $\alpha$-dependent PageRank score over $\alpha$. Importantly, a linear damping can produce results very close to those obtained with PageRank, while requiring fewer iterations to be computed. An important variant of PageRank, EigenTrust, has been proposed to compute trust values in distributed peer-to-peer systems~\cite{KaSchMo03}. EigenTrust replaces uniform teleportation matrix with random jumps to a set of pre-trusted peers, can be easily computed in a distributed way, and is thus suitable for deployment in distributed P2P systems. A very different perspective was adopted in~\cite{PaMaDe11} where a class of quantum PageRank algorithms was proposed based on quantized Markov chains.

Almost at the same time as PageRank, another important algorithm based on random walks and circular reasoning was developed independently. It is called HITS (Hypertext Induced Topic Search) and by contrast to PageRank, it assigns two distinct scores to each node---authority score $x_i$ and hub score $y_i$~\cite{Klein99}. The basic thesis is that a good hub is pointed to by good authorities and vice versa. In mathematical terms, this can be written as
\begin{equation}
\label{HITS}
\vek{x}^{(n+1)}=\mat{A}\T\vek{y}^{(n)},\quad
\vek{y}^{(n+1)}=\mat{A}\vek{x}^{(n+1)}.
\end{equation}
Consequently, one can write $\vek{x}^{(n+1)}=\mat{A}\T\mat{A}\vek{x}^{(n)}$ and $\vek{y}^{(n+1)}=\mat{A}\mat{A}\T\vek{y}^{(n)}$, showing that the stationary authority and hub vectors are the dominant eigenvectors of $\mat{A}\T\mat{A}$ and $\mat{A}\mat{A}\T$, respectively. Since these two matrices are not stochastic matrices as it was the case for PageRank, when finding them by iterations, one has to implement additional normalization of the score vectors. In~\cite{DLK09}, HITS has been generalized to bipartite graphs with the goal to weaken the score reinforcement among the connected nodes and improve the algorithm's robustness to noisy links.
See an extensive review of eigenvector methods for web information retrieval in~\cite{LaMe05}.

In~\cite{CXMR07}, PageRank has been applied to citations among scientific papers (which naturally constitute a directed unweighted network) to assess the relative importance of papers. The authors argued that readers of scientific papers typically follow paths of length two, corresponding to the damping parameter $\alpha=0.5$ much lower than the original value of $0.85$. Albeit the PageRank score of papers was found to be highly correlated with the number of citations (similarly as the PageRank score of web sites is correlated with the number of incoming hyperlinks), significant outliers from this trend were found and identified as seminal publications. This is because the PageRank score redistribution allows a paper with moderate citation count score high thanks to high citation counts of the papers that cite it. As later argued in \cite{MaRe08}, time decay is of crucial importance in the analysis of citation networks because, unlike hyperlinks in the WWW, citations basically cannot be updated after a paper is published. There is also an increasing evidence that time plays an eminent role in the growth of citation networks---see \cite{MeCiGu11} for a recent account. See also~\cite{RaFoVe12} for a general overview of our knowledge about citation networks.

The effect of aging of publications is included in the CiteRank algorithm \cite{WaXiYaMa07} where the uniform teleportation matrix is replaced with $\vek{1}_N\vek{\varrho}\T$ where $\varrho_i=\exp[-t_i/\tau]$, $t_i$ is the age of paper $i$ and $\tau$ is a characteristic decay time. Interestingly, when the correlation between the CiteRank score and the number of recently gained citations is investigated, the optimal damping parameter $\alpha$ is found to be close to the value of 0.5 which was before only hypothesized on the basis of reading habits of researchers. The authors consequently show that apart from selecting papers that contribute most to the current research, CiteRank is particularly successful in selecting papers of long-lasting interest.

Similarly, the network of scientific journals with links weighted by the number of times an article from journal $i$ cites an article from journal $j$ is again suitable for PageRank-like computation of journal status~\cite{BoRoSo06}. Albeit the number of citations does not directly enter here, the resulting ranking of journals is similar to that obtained with the so-called impact factor (which is essentially the average number of citations of recent papers in a given journal). The observed differences in these two measures allowed the authors to introduce the categories of popular journals (which have high impact factors but their citations come from lesser journals, hence the resulting PageRank score of the popular journals is comparatively small) and prestigious journals (which have moderate impact factor but their citations come from journals with high PageRank score, thus allowing the prestigious journals to score high too). A publicly available web site SJR runs a slightly different algorithm based on citations among journals to rank scientific value of journals and countries (see \url{www.scimagojr.com})~\cite{Gon11}.

What is perhaps of even a greater interest to researchers than rankings of papers and journals are rankings of the researchers themselves. The simplest approach to achieve that would be to create a directed networks of authors where links are created according to who cites whom and weight these links according to the citation frequency for a given pair of authors. To better represent the diffusion of scientific credit in such a network, the authors in~\cite{RFMV09} propose additional weights reflecting the number of authors of the citing and cited paper, respectively. If the citing paper $A$ was authored by $n_A$ authors and the cited paper $B$ was authored by $n_B$ authors, $n_An_B$ independent links pointing from an author of paper $A$ to an author of paper $B$ are created, each with weight $1/(n_An_B)$. The credit of individuals is then redistributed over the weighted author-author network in a usual two-fold way: part $1-q$ of $i$'s credit goes to the authors cited by $i$ and part $q$ of $i$'s credit is distributed to all authors according to their productivity. For authors with zero out-strength, it is their whole credit what is distributed to all authors in the network. It is then observed how the resulting ranking of authors changes in time and significant correlations are found between highly-ranked authors and important scientific prizes being given to them. A very similar algorithm has been used to rank professional tennis players~\cite{Rad11}. Another possible approach to the ranking of researchers is by running a PageRank variant on a so-called coauthorship network which is an undirected network where researchers are connected if they have authored a paper together (it is again natural to weight the connection by the number of papers authored together)~\cite{YaDi11}. Co-citation networks where authors are connected if they were cited together by a paper were also used as input for a PageRank-based algorithm to obtain a ranking of authors~\cite{DYFC09}.

PageRank has been used also to measure the importance of species in the network of ecological relationships where the loss of a single species can trigger a cascade of extinctions~\cite{AlPa09}. Upon a minor modification of the input network by introducing a root node which is pointed to by each species and which points back to all ``primary producers'' (species that do not rely on any other species and produce biomass from inorganic compounds) and setting the damping parameter to one (because nutrients cannot randomly jump among nodes in a food web), the authors were able to use the standard PageRank formula. The obtained importance ranking of species was shown to be very effective in choosing nodes leading to the fastest collapse of the food web, outperforming rankings by betweenness and closeness centrality.

A root node pointed by and pointing to all nodes was used also later in~\cite{LZYZ11} where the PageRank algorithm was used to quantify user influence in a directed social network. It is useful to realize that such a root node in fact serves as a teleportation probability: it leads from a given node to the root node and then in the next step to a randomly chosen normal node. This teleportation probability is node-dependent: jump to the ground node occurs with a 50\% probability for a node with only one original out-going link but the probability is only 1\% for a node with 99 original out-going links. In addition, this root node causes the transition matrix to be irreducible and primitive which guarantees existence and uniqueness of a stationary solution. Based on tests on data obtained from the social bookmarking service Delicious.com, the authors of~\cite{LZYZ11} argue that their variant of PageRank is particularly suitable for social networks as it better detects influential users and it is more resistant to manipulations and noisy data.

\subsection{Random walks with sources and sinks}
\label{sec:sosik}
As we have seen above, PageRank is built on a process where the initial node occupancy distribution $\vek{h}^{(0)}$ is gradually washed away by the random walk and an equilibrium distribution $\vek{h}^{(\infty)}$ emerges. In some cases, there exist nodes that act as sources or sinks---they constantly emit or absorb, respectively, ``particles'' that are transported over the network~\cite{StYu07}. To allow for termination of the random walk, it is assumed that sources not only emit new particles but they also absorb particles arriving in them. Denoting the set of source/sink nodes as $S$ and the set of remaining (transient) nodes as $T$ where $\abs{T}:=M$ and thus $\abs{S}=N-M$, we can write the transition matrix in the form
\begin{equation}
\label{trans_m}
\mat{P}=
\begin{pmatrix}
\mat{P}_{SS} & \mat{P}_{ST}\\
\mat{P}_{TS} & \mat{P}_{TT}
\end{pmatrix}
\end{equation}
where we have sorted the nodes so that the first $N-M$ nodes are from $S$ and the next $M$ nodes are from $T$.

If $S$ is the set of sinks, then $\mat{P}_{ST}=\mat{0}$ and $\mat{P}_{SS}=\mat{I}_{N-M}$. We can now ask what is the probability $F_{ij}(t)$ that a particle originating at $i\in T$ gets absorbed in $j\in S$ in $t$ steps or less, avoiding all other sink nodes on its path. This absorption can either occur in one step, with the probability $P_{ij}$, or the particle can first go to another transient node $k$ and then be absorbed from there in $t-1$ steps or less. Together we have
\begin{equation}
\label{F_condition}
F_{ij}(t)=P_{ij}+\sum_{k\in T} P_{ik}F_{kj}(t-1)
\end{equation}
where, of course, $F_{kj}(0)=0$ for all $k$ and $j$. This formula can be written in a matrix form as $\mat{F}(t)=\mat{P}_{TS}+\mat{P}_{TT}\mat{F}(t-1)$ where $\mat{F}(t)$ is an $M\times(N-M)$ matrix of absorption probabilities. The stationary solution $\mat{F}$ thus fulfills $\mat{F}=\mat{P}_{TS}+\mat{P}_{TT}\mat{F}$ and one can express it as
\begin{equation}
\label{Fmat}
\mat{F}=(\mat{I}_M-\mat{P}_{TT})^{-1}\mat{P}_{TS}.
\end{equation}
In the simplest case when $\mat{P}_{TT}=\mat{0}$ (all links from transient nodes lead directly to sink nodes), we obtain $\mat{F}=\mat{P}_{TS}$ as expected. One can show that the inverse $(\mat{I}_N-\mat{P}_{TT})^{-1}$ exists if for every $i\in T$ and $j\in S$ there is a directed path from $i$ to $j$ \cite{StYu07}.

The dual problem of particle diffusion from sources can be solved analogously, leading to the average number of times, $H_{ij}(t)$, that a particle originating at a source node $i$ visits a transient node $j$ in $t$ steps or less, without being absorbed in a source node. The final result reads
\begin{equation}
\label{Hmat}
\mat{H}=\mat{P}_{ST}(\mat{I}_M-\mat{P}_{TT})^{-1}.
\end{equation}
Unlike $\mat{F}$, particle can visit a transient node $j$ repeatedly and therefore $H_{ij}$ can be greater than one. The described picture can be generalized to include the possibility of particle dissipation also in transient nodes~\cite{StYu07}. There is a close relation between random walks with sinks/sources and currents in electric networks---for details see~\cite{DoSn84,Newman05}.

PageRank augmented with sinks was shown to increase the diversity of top ranked items~\cite{LPMWQ10}. After the top ranked object is found by ordinary PageRank computation, it is turned into a sink and the second object is selected from the remaining transient nodes as the one that has the longest time to absorption. The selected node is then turned into a sink too and the third object is again found by the absorption time criterion. Since the expected number of visits of node $j$ when starting in node $i$ is $V_{ij}=[(\mat{I}_M-\mat{P}_{TT})^{-1}]_{ij}$, the expected absorption time of node $i$ is $t_i=\sum_{j} V_{ij}=(\mat{V}\vek{1}_{M})_i$. The absorption time maximization leads to the preference for nodes that are far away in the given network from the nodes already selected for the top of the ranking, which provides a stimulus to the diversity of results.

\begin{figure}[t]
\centering
\onefig{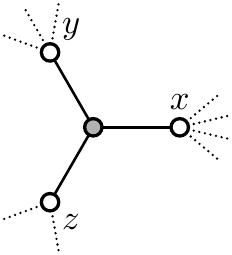}
\caption{In random walk, the occupancy probability of the central node in the next time step is $\tfrac{x}5+\tfrac{y}4+\tfrac{z}3$ (where $x,y,z$ are the current occupancy probabilities of the neighboring nodes, respectively). In heat diffusion, the temperature of the central node in the next time step is $\tfrac{x}3+\tfrac{y}3+\tfrac{z}3$ (where $x,y,z$ are the current temperatures of the neighboring nodes, respectively).}
\label{fig:heat_spreading}
\end{figure}

We finally note a close connection between random walk and heat diffusion. In random walk, the occupancy probability of a node in the next time step depends on the current occupancy probabilities and degrees of its neighbors. By contrast, in heat diffusion the temperature of a node in the next time step depends on the current temperatures of its neighbors and the degree of the given node (see Fig.~\ref{fig:heat_spreading} for an illustration). In mathematical terms, while the transition matrix of random walk reads $P_{ij}=A_{ij}/k_i$ and thus $\mat{P}\T$ is column-normalized, the matrix converting the current vector of temperature values into a next time step vector reads $O_{ij}=A_{ij}/k_j$ and thus $\mat{O}\T$ is row-normalized.

\begin{figure}[t]
\centering
\onefig{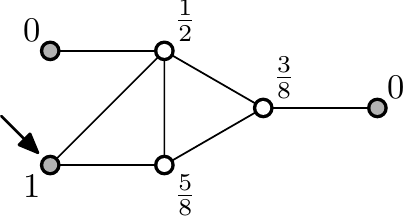}
\caption{Random walk with absorption in sink nodes (shaded with gray): the probability of being absorbed in the arrow-marked node is shown for each node. These probabilities solve the heat equation with the boundary condition given by the temperature of sink nodes fixed at one (for the arrow-marked node) and zero (for all other sink nodes), respectively. For example, the absorption probability $5/8$ for one of the transient nodes can be obtained by averaging the absorption probabilities $1$, $1/2$, and $3/8$ of the neighboring nodes.}
\label{fig:duality}
\end{figure}

Further connections can be found be studying the emission and absorption processes described above. If we fix a sink node $j$, the probabilities of absorption in $j$ for particles starting in node $i$, $F_{ij}$, satisfy the discrete heat equation on the network. This is easy to see on an unweighted undirected network---given a transient node $i$ and its set of neighbors $\mathcal{N}_i$, we can write similarly as in \refeq{F_condition}
$$
F_{ij}=\frac1{k_i}\sum_{k\in\mathcal{N}_i} F_{kj}.
$$
That is, the probability of being absorbed in node $j$ when starting in node $i$ is simply the average over these absorption probabilities when starting in neighbors of node $i$. The boundary condition is given by the sure absorption in $j$ when starting in $j$ and impossible absorption in $j$ when starting in another sink node (corresponding to the boundary probability values $1$ and $0$, respectively). Generalization to a weighted or undirected network is straightforward. This duality is illustrated on a toy network in Fig.~\ref{fig:duality}.

\subsection{Other algorithms}
Node betweenness in a network is calculated as the fraction of the shortest paths between node pairs that pass through a selected node. If the node lies on many shortest paths, it is assumed to be important for information spreading over the network (e.g., it connects extensive clusters). However, node betweenness considers only the shortest paths and thus neglects a significant part of the network's topology. Random-walk betweenness improves this by considering paths of essentially all lengths, albeit still giving more weight to short ones~\cite{Newman05}. It is based on a simple assumption---if random walk starts in node $i$ and ends (gets absorbed) in node $j$, its contribution to the betweenness of node $k$ is given by the average ``net'' number of visits of this node during the random walk, $n_{k}^{(ij)}$. The net number of visits means that passing through a node and then passing through it again from an opposite direction cancel out. Also, if various realizations of random walk are equally likely to pass through a node in opposite directions, these two directions cancel. The resulting betweenness is then obtained by averaging the number of visits over start-end node pairs $(i,j)$
\begin{equation}
b_k=\frac{\sum_{i<j} n_k^{(ij)}}{\tfrac12 N(N-1)}
\end{equation}
where $N$ is the number of nodes in the network. Alternatively, one can obtain the same result building on the electric current injected and removed in a node pair with the contribution to betweenness of node $k$ given by the current passing through this node. The further development is similar to that presented in Sec.~\ref{sec:sosik} and ultimately allows to find betweenness values for all nodes in time $\mathrm{O}((E+N)N^2)$. This betweenness measure is shown to outperform not only the shortest-path betweenness but also the flow betweenness~\cite{Newman05}. With a similar goal, several network flows were typologized and studied by simulations in~\cite{Borg05}.

A very recent second order centrality also makes use of random walks but with three distinctions~\cite{KeMeSeTe11}. Firstly, it can be computed in a distributed manner with nodes having only information of who are their neighbors. Secondly, it relies on ``unbiased'' random walk where the stationary occupancy probability is equal for all nodes regardless of their degree (this is achieved by a Metropolis-Hastings algorithm where step from node $i$ to a neighboring node $j$ is accepted with the probability $k_i/k_j$ for $k_j>k_i$ and always for $k_j\leq k_i$). Finally, it is based on the standard deviation $\sigma_i$ of the return times to a given node $i$. The basic idea is that a node with a central position in the network is visited more regularly than peripheral nodes (those are visited in ``clusters'' with closely grouped subsequent visits interrupted by longer periods when the random walk is in a different part of the network). In addition to numerical stochastic computation of this centrality, various analytical results can be derived and used to better calibrate the numerical implementation.

The network of citations among scientific papers has the special property of being directed and acyclic (the acyclicity is due to citations pointing from a newer paper to older ones). This acyclicity allows one to use the probability of passing through a given node instead of the more traditional occupancy probability. In~\cite{GuMe11}, the probability of passing through node $i$ when the random walk starts in node $j$, $G_{ij}$, was proposed to quantify the influence of node $i$ on node $j$. By summing over $j$, one consequently obtains the aggregate impact of node $i$ which may be in turn used to rank the nodes. Since aggregate impact of node $i$ correlates with the $i$'s progeny size (by $i$'s progeny we mean the set of nodes from which $i$ can be reached by random walk respecting directions of links), one can better distinguish outstanding nodes by comparing the two characteristics. This passing probability framework has been also used to introduce a new node similarity which is based on the assumption that two node's are similar if they are both influenced by the same nodes.

\begin{figure}[t]
\centering
\onefig{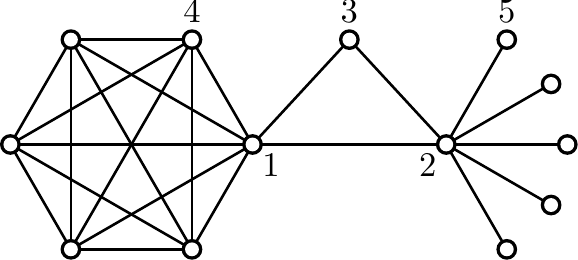}
\caption{A toy network for the computation of node centrality (see results in Tab.~\ref{tab:centrality}).}
\label{fig:centrality}
\end{figure}

To better illustrate performance of the presented methods, we use them to compute node centrality in the network shown in Fig.~\ref{fig:centrality}. (Unlabeled nodes have standing identical with that of node 4 or 5.) For the shortest path centrality (also called betweenness centrality), we count also shortest paths where a node lies on the path's beginning or the path's end. For the PageRank score, we use the usual damping value $\alpha=0.85$. For the random walk centrality, we follow the prescription given in~\cite{Newman05}. For the second order centrality, we convert the standard deviation of return times $\sigma_i$ into a centrality value $1/\sigma_i$ (recall that small $\sigma_i$ is expected for centrally placed nodes). The results summarized in Table~\ref{tab:centrality} show that there are considerable differences between respective centrality measures. While measures agree on a high centrality value of node 1 and a low centrality value of node 5, respectively, big differences exist in assessment of nodes 2, 3, and 4. In particular, eigenvector centrality puts emphasis on the tightly connected part of the network (represented by the complete 6-graph in our toy network) and values little node with low-degree neighbors (in our case, node 2). Random walk centrality awards the central position of node 3 more than other tested measures which is a direct consequence of including not only the shortest paths in computation. One can note that degree centrality and second order centrality rank nodes identically---the value difference between nodes 3 and 4 is however smaller in the case of second order centrality which is again due to its random walk origin being able to appreciate the central location of node 3.

\begin{table}[t]
\centering
\caption{Centrality values for the network shown in Fig.~\ref{fig:centrality}. Values are normalized so that the average centrality is one in all cases.}
\label{tab:centrality}
\setlength{\tabcolsep}{6pt}
\begin{tabular}{rccccc}
\toprule
              & \multicolumn{5}{c}{node}\\
                \cmidrule(lr){2-6}
      measure & 1 & 2 & 3 & 4 & 5\\
\midrule
       degree & 1.98 & 1.98 & 0.57 & 1.41 & 0.28\\
shortest path & 2.59 & 3.14 & 0.66 & 0.66 & 0.66\\
  eigenvector & 2.03 & 0.62 & 0.52 & 1.84 & 0.12\\
     PageRank & 1.71 & 2.65 & 0.68 & 1.12 & 0.47\\
  random walk & 2.31 & 2.69 & 1.09 & 0.84 & 0.55\\
 second order & 2.23 & 2.23 & 0.87 & 1.17 & 0.36\\
\bottomrule
\end{tabular}
\end{table}

\section{Recommendation}
\label{sec:recommendation}
The task of recommender systems is to utilize past evaluations of items by users to select further items that could be appreciated by the users. We often speak about personalized recommendations because a good recommender system should be able to recognize preferences of individuals and select the object to be recommended accordingly. Thanks to the availability of large-scale data on user behavior and the ever-increasing power of computers at our disposal, the field of recommendation grows rapidly. Nowadays, one can hardly imagine a successful e-commerce site without a sophisticated recommender system (think of Amazon.com) and companies organize competitions aiming to improve their recommendation methods (as it was prominently done by Netflix by their NetflixPrize)~\cite{SKR01}. Approaches used to produce recommendations range from variants of the simple thesis ``recommend to a user what was already appreciated by similar users'' to complicated  mathematical models and machine learning techniques~\cite{AdoTuz2005,Ricci2011,LMYZZZ12}. The problem of link prediction is closely related to recommendation with the task being to identify possible missing or future links in a given network~\cite{LuZh11}.

In this section, we aim to discuss the use of random walks in recommendation. First of all, similarity measures based on random walks can be used in similarity-based (sometimes called memory-based) collaborative filtering algorithms. Denoting the rating of object $\alpha$ given by user $i$ as $r_{i\alpha}$ and the average rating of user $i$ as $\mu_{i}$, the generic form of collaborative filtering using user similarity is
\begin{equation}
\tilde r_{i\alpha}=\mu_i+\frac{\sum_{j\in R_{\alpha}} s_{ij}(r_{j\alpha}-\mu_j)}
{\sum_{j\in R_{\alpha}} \abs{s_{ij}}}
\end{equation}
where $\tilde r_{i\alpha}$ is the expected (predicted) rating of object $\alpha$ by user $i$ and $R_{\alpha}$ is the set of users who have already rated object $\alpha$. User similarity $s_{ij}$ (or, object similarity $s_{\alpha\beta}$ for an item-based variant of collaborative filtering) is usually computed using the standard Pearson similarity or cosine similarity. Our interest now is in random walk-based similarity measures that can be used instead of traditional ones.

Assuming that random walk starts in node $i$, one can introduce the average first passage time for node $j$, $T(j\vert i)$. The symmetrized quantity $C(i,j):=T(j\vert i)+T(i\vert j)$, the average commute time, was shown to act as a distance on the graph and can be further transformed into $\sqrt{C(i,j)}$, a so-called Euclidean Commute Time Distance~\cite{FoPiReSa07}. In addition, both $C(i,j)$ and $\sqrt{C(i,j)}$ can serve as node similarity measures and in turn effectively used for collaborative filtering. While one can compute $C(i,j)$ on a node-by-node basis using the sink-node machinery described in Sec.~\ref{sec:sosik}, it is computationally more efficient to employ the formula
\begin{equation}
C(i,j)=2E\big(l_{ii}^{+}+l_{jj}^{+}-2l_{ij}^{+}\big)
\end{equation}
where $l_{ij}^{+}$ is an element of the Moore-Penrose pseudoinverse $\mat{L}^+$ of the network's Laplacian matrix $\mat{L}=\mat{D}-\mat{A}$ (here $\mat{D}$ is the degree matrix with elements $d_{ij}=k_i\delta_{ij}$)~\cite{FoPiReSa07}. Pseudoinverse is applied because $\mat{L}$ cannot be inverted (zero is one of its eigenvalues) and can be computed as $\mat{L}^+=(\mat{L}-\vek{1}_N\vek{1}_N\T/N)^{-1}+\vek{1}_N\vek{1}_N\T/N$.

A simple node similarity measure based on local random walk was proposed in~\cite{LiLu10}. Denoting the probability that a random walker starting at node $i$ is located at node $j$ after $t$ time steps as $\pi_{ij}(t)$, the similarity of nodes $i$ and $j$ was proposed in the form
\begin{equation}
s_{ij}^{LRW}(t)=\frac1{2E}\big(k_i\pi_{ij}(t)+k_j\pi_{ji}(t)\big)
\end{equation}
where $E$ is the total number of edges in the graph. Multiplication with node degree ($k_i$ and $k_j$, respectively) gives more weight to nodes with high degree and compensates for the increased dispersion of random walk at those nodes (if many links lead from $x$, $\pi_{xy}$ can be low). The obtained quantity can be summed over different $t$, leading to ``superposed'' similarity $s_{ij}^{SRW}(t)=\sum_{\theta=1}^t s_{ij}^{RW}(\theta)$. Numerical evaluation on five distinct real networks showed that $s^{LRW}$ and $s^{SRW}$ in most cases outperform traditional similarity metrics in accuracy and are less computationally demanding than other well performing methods~\cite{LiLu10}. A method for random walk computation of paper similarity was proposed specifically for scientific citation data~\cite{GuYe11}. When computing similarity of papers $i$ and $j$, two two-step random walks are combined. One aims ``downstream'' to papers cited by both $i$ and $j$, thus reflecting the opinion of the authors of $i$ and $j$. The other aims ``upstream'' to papers citing both $i$ and $j$, thus reflecting the opinion of the readers of $i$ and $j$. It is then shown that this novel similarity measure is able to identify the backbone of the citation network, leading to accurate characterization of hierarchical structure of the scientific development and its classification into fields and sub-fields.

Due to sparsity of the input data, traditional similarity measures based on overlapping neighborhoods can fail to accurately assess node similarity. To alleviate this problem, it was suggested to transform the similarity matrix into a PageRank-like form $\mat{P}$ by normalization and addition of random jumps, and then use $\mat{P}(\mat{1}-\alpha\mat{P})^{-1}$ as a new similarity matrix where similarity values are assigned also to item pairs that have not been evaluated by any users~\cite{YiKr08}. Here $\alpha\in(0,1)$ is the probability of continuing the random walk and thus $1/\alpha$ is the characteristic number of steps over which similarity is transferred.

\begin{figure}[t]
\centering
\onefig{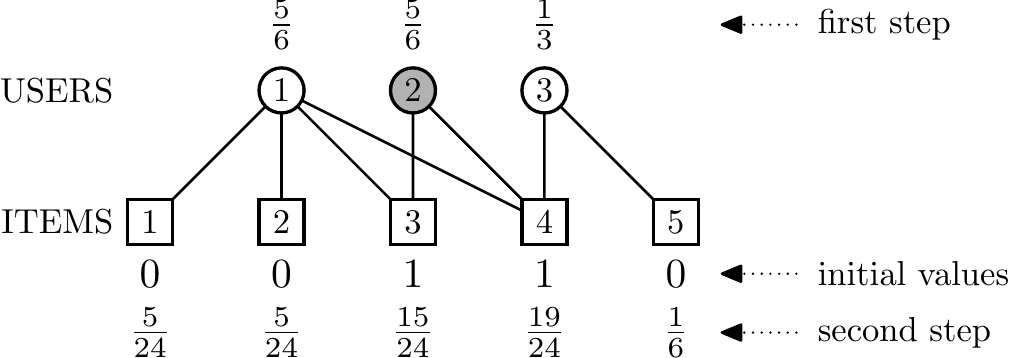}
\caption{Illustration of random walk recommendation for user 2. Items collected by user 2 are initially assigned unit resource which then spreads uniformly to users connected with these items and finally back to the item side. Items with the highest resulting resource amount are then recommended to the given user. In this case, items 1 and 2 score best (items 3 and 4 have higher resulting values but are ignored as they have been already collected by user 2).}
\label{fig:ProbS}
\end{figure}

Apart from using random walks to quantify node similarity, there are also recommendation methods that are directly based on random walks. In~\cite{Zhou07}, the authors consider the bipartite user-item network where links connect users with the items they collected or appreciated. Note that explicit ratings given by users to items play no role here---the method only requires the knowledge of items that have been collected/favored by individual users. Assuming that each item collected by a given user $i$ is assigned a unit initial resource, this resource is spread uniformly from the collected items to the users connected with them and then in the second step back to items connected with those users (see Fig.~\ref{fig:ProbS} for an illustration). The final amount of resource on respective items is then interpreted as their recommendation score and items with the highest score are then recommended to user $i$ (already collected items are of course excluded). The reasoning behind this spreading process is that it selects items that have been collected by users who share some interests with the given user $i$. At the same time, if user $i$ has collected an extremely popular item $\alpha$ or if a collected item has been co-collected by an extremely active user $j$, the information signal is weak because the overlap between $i$ and $j$ as well as between $i$ and $\alpha$ is rather small in those cases. The random walk-based even spreading of the resource is thus a reasonable approach to quantify the resulting recommendation scores.

The transition matrices from objects to users and vice versa have the form $U_{i\alpha}=A_{i\alpha}/k_{\alpha}$ and $V_{\alpha i}=A_{i\alpha}/k_i$, respectively, where $k_{\alpha}$ is the degree of item $\alpha$ (the number of users who collected it) and $k_i$ is the degree of user $i$ (the number of items collected by this user). The vector with object recommendation scores for user $i$ then reads $\tilde{\vek{h}}_i=\mat{V}\mat{U}\vek{h}_i$ where $(\vek{h}_i)_{\alpha}=A_{i\alpha}$ encodes which items have been actually collected by user $i$. One can introduce $\mat{W}^P:=\mat{V}\mat{U}$ and show that
\begin{equation}
\label{ProbS}
W_{\alpha\beta}^P=\frac1{k_{\beta}}\sum_{i=1}^U\frac{A_{i\alpha}A_{i\beta}}{k_i}
\end{equation}
where indices $\alpha$ and $\beta$ are used to enumerate items, $i$ enumerates users and $U$ is the total number of user nodes. One can also spread the initial resource over $2n$ steps in the bipartite network by $(\mat{W^P})^n\vek{h}_i$ but the result converges fast to a vector whose elements are proportional to object degree $k_{\alpha}$ and hence conveys little information for personalized recommendation.

This basic method has been subsequently generalized in multiple ways. For example, it was proposed to assign the initial amount of resource to items not uniformly but depending on the item degree as $k_{\alpha}^{\theta}$~\cite{Zhou2008}. Best results were achieved with $\theta\approx -1$ when the produced recommendations were both more accurate and more personalized. To better answer the need for diversity in recommendations, a hybrid algorithm was proposed which combines the random walk algorithm with heat spreading~\cite{Zhou_PNAS}. As we have already seen, heat diffusion differs from random walk in normalization of their matrices and thus the matrix of heat diffusion reads $W_{\alpha\beta}^H=(1/k_{\alpha})\sum_{i=1}^U A_{i\alpha}A_{i\beta}/k_i$. The best performing hybrid of the two has the form
\begin{equation}
\label{hybrid}
W_{\alpha\beta}^{P+H}=\frac1{k_{\alpha}^{1-\lambda}k_{\beta}^{\lambda}}
\sum_{i=1}^U\frac{A_{i\alpha}A_{i\beta}}{k_i}
\end{equation}
where the parameter $\lambda$ controls the balance between the contribution of random walk and heat spreading. This method was shown to simultaneously increase accuracy and diversity of recommendations.

A combination of random walk and heat diffusion for data with explicit ratings was presented in~\cite{Blattner10} where recommendation scores obtained by each respective process are multiplied to obtain the final recommendation score. In addition, the employed random walk is self-avoiding, i.e., there is no possibility to return to the initial item node after two steps. If user evaluations are given in an integer scale (a very typical case nowadays), a multichannel spreading can be employed where the states of the random walk are represented not only by the current item but also by the rating given to this item~\cite{ZMRZLY07}. If, for example, a five-level rating scale is used, $5\times5$ connections are created between any two items. However, this approach suffers from aggravating the data sparsity problem (the same amount of data is used to construct many more connections between (item,rating) pairs) which limits its performance.

Spreading over a bipartite network is considered also in~\cite{SiGuMeSu07} where the bipartite user-item network is augmented with social links among users (this kind of data is often produced in online gaming). The random walk starting at the user for which recommendations are being made follows a social link to another user with probability $\alpha$ or a link to an item with probability $1-\alpha$ where it is absorbed. Items are then ranked according to the fraction of random walks absorbed in them. A different mechanism of heat diffusion on an item-item network was used to produce recommendations by representing items liked and disliked by a given user as nodes with fixed temperature $1$ and $0$, respectively~\cite{ZhBlYu07}. From the remaining nodes, those with the highest resulting temperature are then chosen to be recommended to the given user. See \cite[Ch. 6]{LMYZZZ12} for other related works and more detailed information.

\section{Conclusion}
We attempted here to give an overview of applications of random walks to information filtering, focusing on the tasks of ranking and recommendation in particular. Despite the amount of work done in these two directions, multiple important research challenges still remain open. Due to the massive amounts of available data, scalability of algorithms is of critical importance. Even when full computation is possible, one can think of potential approaches to update the output gradually when new data arrives. To achieve that, one can use or learn from perturbation theory which is a well-known tool in physics. We have seen that results based on random walks often correlate strongly with mere popularity (represented by degree) of nodes in the network. Such bias toward popularity may be beneficial for an algorithm's accuracy but it may also narrow our view of the given system and perhaps create a self-reinforcing loop further boosting popularity of already popular nodes. We thus need information filtering algorithms that converge less to the center of the given network. Random walks biased by node centrality or time information about nodes and links could provide a solution to this problem. As a beneficial side effect, this line of research could yield algorithms pointing us to fresh and promising content instead of highlighting old victors over and over again.

\section{Acknowledgments}
This work was partially supported by the Swiss National Science Foundation Grant No. 200020-132253. I wish to thank a number of wonderful friends and colleagues who helped to shape many of the ideas presented here.

\printindex
\end{document}